\documentclass[twocolumn,showpacs,preprintnumbers,amsmath,amssymb,superscriptaddress]{revtex4}

\usepackage{graphicx}
\usepackage{dcolumn}
\usepackage{bm}
\usepackage{braket}

\begin{document}


\title{NMR Investigation of the Quasi One-dimensional Superconductor K$_{2}$Cr$_{3}$As$_{3}$}

\author{H.\ Z.\  Zhi}
\affiliation{Department of Physics and Astronomy, McMaster University, Hamilton, Ontario L8S4M1, Canada} 
\author{T.\  Imai}\email{imai@mcmaster.ca}%
\affiliation{Department of Physics and Astronomy, McMaster University, Hamilton, Ontario L8S4M1, Canada} 
\affiliation{Canadian Institute for Advanced Research, Toronto, Ontario M5G1Z8, Canada} 
\author{F.\ L.\ Ning}
\affiliation{Department of Physics, Zhejiang University, Hangzhou 310027, China} 
\affiliation{Collaborative Innovation Centre of Advanced Microstructures, Nanjing University, Nanjing 210093, China}
\author{Jin-Ke Bao}
\affiliation{Department of Physics, Zhejiang University, Hangzhou 310027, China} 
\affiliation{Collaborative Innovation Centre of Advanced Microstructures, Nanjing University, Nanjing 210093, China}
\author{Guang-Han Cao}
\affiliation{Department of Physics, Zhejiang University, Hangzhou 310027, China} 
\affiliation{Collaborative Innovation Centre of Advanced Microstructures, Nanjing University, Nanjing 210093, China}

\date{\today}

\begin{abstract}
We report $^{75}$As NMR measurements on the new quasi one-dimensional superconductor K$_{2}$Cr$_{3}$As$_{3}$ ($T_{c} \sim 6.1$~K) [J.\ K.\ Bao et al., Phys. Rev. X {\bf 5}, 011013 (2015)].  We found evidence for strong enhancement of Cr spin fluctuations above $T_c$ in the [Cr$_{3}$As$_{3}$]$_{\infty}$ double-walled subnano-tubes based on the nuclear spin-lattice relaxation rate $1/T_{1}$.  The power law temperature dependence, $1/T_{1}T \sim T^{-\gamma}$ ($\gamma \sim 0.25$), is consistent with the Tomonaga-Luttinger liquid.  Moreover, absence of the Hebel-Slichter coherence peak of $1/T_{1}$ just below $T_{c}$ suggests unconventional nature of superconductivity.
\end{abstract}

\pacs{74.70.Xa, 76.60.-k, 71.10.Pm}
\maketitle  
The surprising discovery of high $T_c$ superconductivity in an iron-pnictide LaFeAsO$_{1-x}$F$_{x}$ \cite{Kamihara} led to frantic search for superconductivity in iron-based pnictides and chalcogenides \cite{Review}.  The key building unit of these iron-based high $T_c$ superconductors is the square lattice formed by Fe atoms.  The mechanism of high $T_c$ superconductivity in iron-pnictides remains as controversial as that in copper-oxides, which is also comprised of the square lattice of Cu atoms.  The recent discovery of superconductivity in helimagnetic CrAs with $T_{c} \sim 2.2$\ K under modest applied pressure of $\sim 0.7$\ GPa \cite{Wu, Kotegawa} expanded the  avenue of superconductivity research to Cr-pnictides.  

Very recently, Bao et al. discovered superconductivity in K$_{2}$Cr$_{3}$As$_{3}$ with the onset of $T_{c} = 6.1$\ K in ambient pressure \cite{Bao}.  Replacing K$^{+}$ ions with larger Rb$^{+}$ and Cs$^{+}$ ions lowers the $T_c$ to 4.8~K \cite{Tang} and 2.2~K \cite{Tang2}, respectively.  In Fig.\ 1, we present the proposed crystal structure of K$_{2}$Cr$_{3}$As$_{3}$ with the most probable space group $\overline{P}$6m2 (No. 187) \cite{Bao}.  The key building unit of K$_{2}$Cr$_{3}$As$_{3}$ is the one dimensional [(Cr$_{3}$As$_{3}$)$^{2-}$]$_{\infty}$ double-walled subnano-tubes (DWSTs) with the outer diameter of 0.58 nm separated by columns of K$^{+}$ ions, in striking contrast with the two dimensional square lattice that forms iron-pnictide and copper-oxide high $T_c$ superconductors.  Assuming the standard ionic valence As$^{3-}$ and ignoring vacancies in the lattice, the nominal valence of the transition metal element is Cr$^{2.3+}$, and hence each Cr has 3.7 3d electrons on average.  The short interatomic distance between Cr atoms, 2.61 to 2.69 \AA, suggests that the Cr-Cr bonding is metallic, while the bonding of Cr with As$^{3-}$ ions may be considered ionic \cite{Bao}.  The bulk physical property measurements on the critical field $H_{c2}$ \cite{Bao}, the electronic specific \cite{Bao, Tang}, and the penetration depth \cite{Yuan} point toward the presence of a node(s) in the superconducting energy gap.

\begin{figure}
\includegraphics[width=3in]{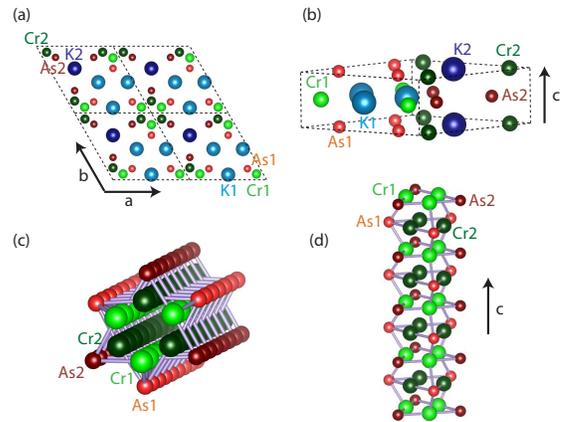}
\caption{\label{fig:Structure} (Color online) The crystal structure of quasi 1D superconductor K$_{2}$Cr$_{3}$As$_{3}$ \cite{Bao}.  (a) Top view of four unit cells.  A complete [Cr$_{3}$As$_{3}$]$_{\infty}$ DWST is in the middle.  (b) Angled view of a unit cell, and (c)-(d) DWST.}
\end{figure}

Besides the generic interest in the mechanism of exotic superconductors, the novel linear chain structure of the [Cr$_{3}$As$_{3}$]$_{\infty}$ DWSTs provides us with a unique opportunity to investigate the fundamental physics of a quasi one dimensional (1D) inorganic metal, and its relation with superconductivity \cite{Bao}.  Theoretically, it is well established that Fermions confined in 1D, commonly called the Tomonaga-Luttinger liquid (TLL), behave very differently from the two or three dimensional analogues, because electron-electron interactions completely alter the electronic properties near the Fermi energy no matter how weak the interactions are \cite{Tomonaga, Luttinger, Lieb}.  As a consequence, a simple Fermi liquid theory based on Landau's quasi-particle picture breaks down.  The peculiar influence of interactions between two particles in the TLL could be understood intuitively, if we realize that two particles must always meet each other head on in 1D; they cannot avoid each other by moving side ways.  A fingerprint of the TLL is the power-law behavior arising from the singularity at the Fermi energy that manifest in various physical properties.  Past explorations of the exotic properties of the TLL focused on organic conductors (see, for example, discussions in \cite{Bourbonnais, Bourbonnais2, Organic} and references therein), semiconductor nano structures \cite{Semicon}, carbon nanotubes \cite{Nature, Yoshioka, Singer, German, Ihara} and their superconductivity \cite{Carbon1, Carbon2}, or quasi-1D Heisenberg model systems \cite{Klanjsek, Kuhne1, Kuhne2}.  Does the [Cr$_{3}$As$_{3}$]$_{\infty}$ DWST in K$_{2}$Cr$_{3}$As$_{3}$ indeed exhibit the signatures expected for the TLL above $T_c$?  If so, is the superconducting state below $T_c$ also exotic?      

In this paper, we report the first microscopic $^{75}$As NMR investigation of K$_{2}$Cr$_{3}$As$_{3}$ by fully taking advantage of the versatile nature of the NMR techniques.  We probed the electronic and superconducting properties at different $^{75}$As sites separately by measuring their nuclear spin-lattice relaxation rate $1/T_1$.  We found evidence for strong enhancement of Cr spin fluctuations toward $T_c$.  The dynamical electron spin susceptibility $\chi''$ of the DWST obeys a characteristic temperature dependence, $\chi'' \propto 1/T_{1}T \sim T^{-\gamma}$ ($\gamma = 0.25 \pm 0.03$).  The observed power law behavior resembles that of the organic conductor TTF[Ni(dmit)$_{2}$]$_{2}$ \cite{Bourbonnais2} and carbon nanotubes \cite{Yoshioka, Singer, German, Ihara}, and is consistent with the TLL.  Moreover, we show that the Hebel-Slichter coherence peak of $1/T_1$ \cite{Hebel}, commonly observed for conventional BCS s-wave superconductors with isotropic energy gaps, is absent in the present case.  

In view of the low symmetry of the $^{75}$As sites in K$_{2}$Cr$_{3}$As$_{3}$, the EFG (electric field gradient) at the $^{75}$As sites originating from K$^{+}$, Cr$^{2.3+}$, and As$^{3-}$ ions in their vicinity must be large.  Since the nuclear quadrupole interaction frequency $\nu_{Q}$ of the $^{75}$As nuclear spins ($I = 3/2$, nuclear gyromagnetic ratio $\gamma_{n}/2\pi = 7.2919$\ MHz/T) is proportional to the EFG, we anticipated that $\nu_{Q}$ in K$_{2}$Cr$_{3}$As$_{3}$ should be much larger than $\nu_{Q} \sim 2$~MHz in the iron-pnictide high $T_c$ superconductor Ba(Fe$_{1-x}$Co$_{x}$)$_{2}$As$_{2}$ \cite{NingJPSJ1}.  (The $^{75}$As site in Ba(Fe$_{1-x}$Co$_{x}$)$_{2}$As$_{2}$ is more symmetrical and surrounded by Ba$^{2+}$ and Fe$^{2+}$ ions.)  We first searched for the $^{75}$As NMR signals in our randomly oriented powder sample in high magnetic fields, and identified two distinct $^{75}$As NMR signals with $\nu_{Q} \sim 40$\ MHz \cite{Supplement}.  Such large values of $\nu_{Q}$ would readily allow us to detect $^{75}$As NQR (Nuclear Quadrupole Resonance) between the (nominal) $I_{z} = \pm 3/2$ and $\pm1/2$ transitions in zero external magnetic field, $B = 0$.  NQR is advantageous in probing the intrinsic superconducting properties, because we do not perturb the superconducting state with the applied magnetic field.  

Armed with the preliminary knowledge of $\nu_{Q}$, we searched for the $^{75}$As NQR signals between 33 and 55 MHz.  We show representative $^{75}$As NQR spectra in Fig.\ 2.  The lineshape observed at 200~K indicates the presence of two sets of sharp NQR peaks: the A line near 39\ MHz and the B line near 44\ MHz, both accompanied by smaller side peaks.  Since the integrated intensity of the NQR signals below 43~MHz is equal to that above 43~MHz, these two sets of NQR signals must arise from As1 and As2 sites.  Notice that the local arrangements of K$^{+}$ ions near As1 and As2 sites are different, as readily seen in Fig.1(a,b).  Therefore $\nu_{Q}$ should be somewhat different, too, between As1 and As2 sites.  Unless we conduct single crystal NMR measurements, we are unable to determine which of the A and B lines arise from As1 and As2 sites, but none of our discussions below depend on the details of the site assignments.  

The presence of smaller side peaks and broad continuum suggests the influence of K$^{+}$ defects on $\nu_{Q}$, as often observed in alloys and disordered materials.  According to the energy-dispersive X-ray spectroscopy (EDS) analysis, the composition of the present material is actually close to K$_{1.82\pm 0.19}$Cr$_{3}$As$_{2.99\pm0.07}$ \cite{Bao}.  Generally, defects would locally alter the magnitude of $\nu_{Q}$ through the change of the EFG tensor (see \cite{Takeda} for a recent example of Cu substitution effects on $\nu_{Q}$ in the BaFe$_{2}$As$_{2}$ high $T_c$ superconductor).  Since the side peak of the B line is separately observable even at low temperatures, we distinguish the main and side peaks by calling them B$_{1}$ and B$_{2}$, respectively.  The temperature dependence of the NQR frequency $\nu_{Q}$ is very similar at A, B$_{1}$ and B$_{2}$ sites as shown in the inset to Fig.\ 2.  The decrease of $\nu_{Q}$ from 2\ K to 295\ K is anomalously large, $3 \sim 5$~\%.  The absence of a kink or jump in the temperature dependence of $\nu_{Q}$, however, rules out the presence of Peierls instability.   

\begin{figure}
\includegraphics[width=3.2in]{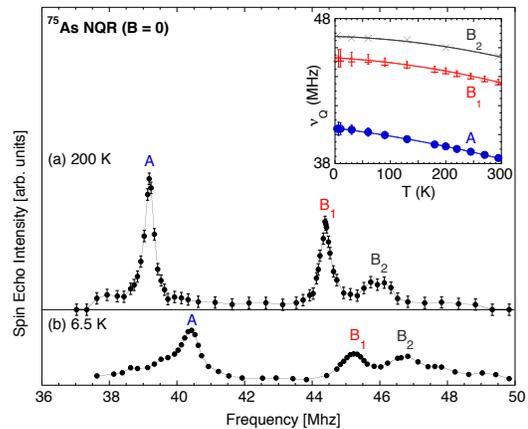}
\caption{\label{fig:Structure} (Color online) $^{75}$As NQR spectra measured at (a) 200\ K and (b) 6.5\ K.  We divided the raw NQR signal intensity by  the square of the frequency to take into account variation of the sensitivity.  We marked three distinctive peaks as A, B$_{1}$, and B$_{2}$.  Inset: the temperature dependence of the $^{75}$As NQR peak frequency, $\nu_{Q}$.  All solid curves are guides for the eye.}
\end{figure}

\begin{figure}
\includegraphics[width=2.8in]{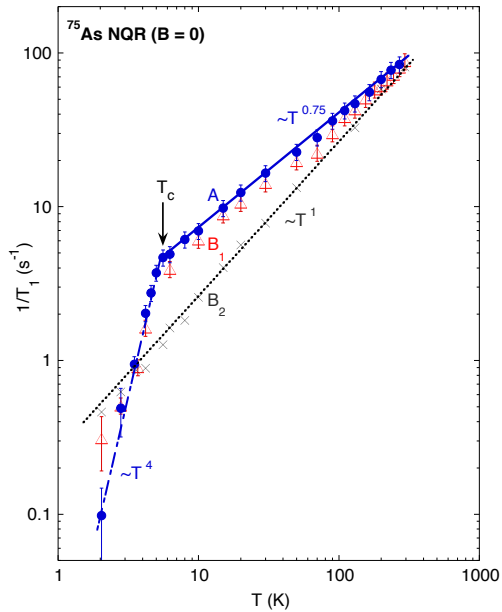}
\caption{\label{fig:Structure} (Color online) $^{75}$As nuclear spin-lattice relaxation rate $1/T_{1}$ measured by NQR techniques in $B=0$ for A, B$_{1}$, and B$_{2}$ sites.  The solid line above $T_c$ is the best fit for the A sites with a power-law, $1/T_{1} \sim T^{1-\gamma}$ ($\gamma = 0.25 \pm 0.03$), while the dashed-dotted line below $T_c$ represents $1/T_{1} \sim T^{4}$.  The dotted line is the best linear fit of the normal state data for the B$_{2}$ peak, $1/T_{1} = 0.27\cdot T$\ s$^{-1}$.  
}
\end{figure}

\begin{figure}[t]
\includegraphics[width=3.2in]{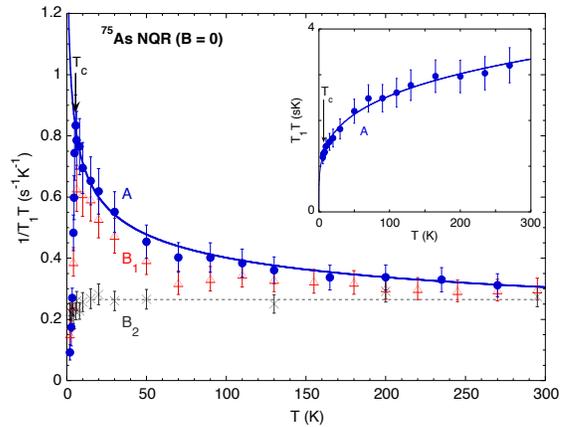}
\caption{\label{fig:Structure} (Color online) $1/T_{1}T$ for A, B$_{1}$, and B$_{2}$ sites.  The solid curve through the normal state data of the A sites represents the power-law fit, $1/T_{1}T \sim T^{-\gamma}$, with the same $\gamma$ as in Fig.\ 3.    The dotted straight line through the data points of the B$_{2}$ peak is the Korringa fit, $1/T_{1}T = 0.27$\ s$^{-1}$K$^{-1}$ above $T_c$.   Inset: Temperature dependence of $T_{1}T$ for the A peak above $T_c$, with the best fit, $T_{1}T \sim T^{+\gamma}$. }
\end{figure}

Next, let us examine the electronic properties of the [(Cr$_{3}$As$_{3}$)$^{2-}$]$_{\infty}$ DWSTs.  In Fig.\ 3, we plot the temperature dependence of $1/T_{1}$ measured by inversion recovery techniques \cite{Supplement}  at $^{75}$As sites in a log-log scale.  We also present $1/T_{1}T$ in Fig.\ 4.  Quite generally, $1/T_{1}T$ probes the wave-vector {\bf q}-integral in the first Brillouin zone of the imaginary part of the dynamical electron spin susceptibility, $\chi''({\bf q}, \nu_{Q})$, where $\nu_{Q}$ is the NQR frequency used to measure $1/T_1$.  In other words, $1/T_{1}T$ probes the low frequency spin dynamics of electrons integrated over the Brillouin zone.  If the underlying electronic states of the DWSTs may be described by a simple Fermi liquid theory and the electron-hole pair excitations dominated the low energy excitations at the Fermi surface, we expect $1/T_{1} \propto T \cdot N(E_{F})^2$, where $N(E_{F})$ is the electronic density of states at the Fermi energy $E_F$.  That is, Korringa relation holds, $1/T_{1}T \sim$ constant, for Fermi liquids with a broad band(s).  The strong increase of $1/T_{1}T$ toward $T_c$ observed for the main A and B$_{1}$ sites clearly contradicts with such an expectation.  Without relying on any theoretical assumptions, we conclude that Cr spin fluctuations grow toward $T_c$ prior to the onset of superconductivity at a certain wave vector(s).  

The magnitude of $1/T_{1}T$ in Fig.\ 4 is comparable to that of the iron-pnictide high $T_c$ superconductors such as Ba(Fe$_{1-x}$Co$_{x}$)$_2$As$_2$ \cite{Ning, Nakai, Zheng}.  If we assume that the hyperfine coupling of $^{75}$As nuclear spins with Cr 3d electron spins in the present case is comparable to that with Fe 3d electron spins in iron-pnictides, our results in Fig.\ 4 imply that the dynamical spin susceptibility of Cr is enhanced near $T_c$ as much as the case of the superconducting Ba(Fe$_{0.92}$Co$_{0.08}$)$_2$As$_2$ near its $T_{c} \sim 25$\ K \cite{Ning}.  At first glance, the gradual growth of  $1/T_{1}T$ (and hence $\chi''$) toward $T_c$ in the present case also seems similar to typical iron-pnictide high $T_c$ superconductors \cite{Ning, Nakai, Zheng}.  The growth of $\chi''$ near $T_c$ within the FeAs planes of iron-pnicides could be successfully fit with a Curie-Weiss law, $1/T_{1}T \sim C/(T+\theta)$, where $C$ and $\theta$ are constants, in the temperature range where the electrical resisitivity shows a linear temperature dependence \cite{Ning, Nakai, Zheng}.  The Curie-Weiss behavior is theoretically anticipated for two-dimensional electron gas systems with antiferromagnetic spin correlations \cite{Millis, Moriya}.  

In contrast with the case of quasi two-dimensional iron-pnictide high $T_c$ superconductors, the fundamental structural unit of quasi 1D K$_{2}$Cr$_{3}$As$_{3}$ is a metallic DWST formed by [Cr$_{3}$As$_{3}$]$_{\infty}$.  Recent theoretical calculations predicted the existence of two quasi 1D bands and one three dimensional band \cite{Band, WuHu}, and very strong antiferromagnetic correlations along the DWST with the nearest-neighbor Cr-Cr exchange interaction as large as $\sim 1000$~K \cite{WuHu}.  We also point out that the DWST's form a triangular-lattice, as shown in Fig.\ 1(a).  This means that three-dimensional antiferromagnetic couplings between the adjacent DWST's are geometrically frustrated.  The quasi 1D nature of spin correlations is therefore protected, suggesting the predominantly 1D character of Cr spin fluctuations enhanced near $T_c$. 

As explained above, the 1D electron gas with electron-electron interactions forms a TLL that gives rise to a power law behavior in various physical observables.  Earlier theoretical calculations showed that the TLL would show a power-law behavior, $1/T_{1}T \sim T^{-\gamma}$, where the exponent $\gamma$ is a non-universal constant that depends on the details of the system, such as the band structure, nesting wave vector $2k_{F}$, and the strength of interactions \cite{Bourbonnais, Yoshioka, German}.  Analogous power-law behavior was previously reported for TTF[Ni(dmit)$_{2}$]$_{2}$ with $\gamma \sim 0.7$ \cite{Bourbonnais2} and for single-wall carbon nanotubes with $\gamma \sim 0.66$ \cite{Ihara}.  

Close examination of our $1/T_{1}$ data in Fig.\ 3 indeed reveals that all of our $1/T_{1}$ data points for the A (and B$_1$) sites above $T_c$ are on a straight line in a log-log plot, implying a power-law, $1/T_{1} \sim T^{1-\gamma}$, or equivalently, $\chi'' \propto 1/T_{1}T \sim T^{-\gamma}$.  The best fit yields $\gamma = 0.25 \pm 0.03$.  We overlay a corresponding power-law fit with the same $\gamma$ in Fig.\ 4, which nicely reproduces the mysteriously strong divergent behavior of $\chi''$ near $T_c$.  Our findings are similar to earlier reports on quasi 1D materials with the TLL behavior \cite{Bourbonnais, Bourbonnais2, Organic, Singer, German, Ihara, Kuhne1, Kuhne2}.  The observed value of $\gamma = 0.25$ in the present case  suggests that the electron-electron interactions are repulsive, and the dominant channel of the spin correlations is antiferromagnetic for the wave vector $2k_{F}$ \cite{Bourbonnais}.  Our conclusion is consistent with the large antiferromagnetic exchange coupling $\sim 1000$~K expected for nearest-neighbor Cr-Cr spins along the DWST \cite{WuHu}.  We call for additional experimental tests of the TLL behavior of the DWSTs in K$_{2}$Cr$_{3}$As$_{3}$ based on ARPES and other techniques.                     
 
Having established the characteristic quasi 1D behavior of Cr spin fluctuations above $T_c$, let us turn our attention to the superconducting state.  In conventional isotropic BCS s-wave superconductors, $1/T_{1}$ exhibits a hump just below $T_c$ due to the sharp density of states at the edge of the energy gap, where the low energy quasi-particle excitations contribute constructively to $1/T_{1}$ due to the coherence factor predicted by the BCS theory \cite{Hebel, MacLaughlin}.  The observation of such a Hebel-Slichter coherence peak of $1/T_1$ is a crucial test for the validity of the description of the superconducting state based on the conventional isotropic BCS s-wave model.  In addition, $1/T_{1}$ decreases exponentially far below $T_c$, $1/T_{1} \sim exp(-\Delta/k_{B}T_{c})$, in isotropic BCS s-wave superconductors, where $\Delta$ is the isotropic energy gap at the Fermi surface \cite{Masuda, MacLaughlin}.  For contemporary examples of the conventional BCS s-wave superconductors, see \cite{MgB2, SingerMgCNi3}.

In the present case, our $1/T_{1}$ data for the A and B$_1$ sites in Fig.\ 3 show a steep drop just below $T_{c}$\ without exhibiting the coherence peak.  Moreover, the temperature dependence of $1/T_{1}$ below $T_c$ is consistent with a power law, $1/T_{1} \sim T^{n}$ with $n \sim 4$.  These results below $T_c$ are similar to the case of unconventional superconductors, such as the high $T_c$ superconductor YBa$_2$Cu$_3$O$_{7-\delta}$ with d-wave pairing symmetry \cite{Imai}.  We are not aware of any theoretical prediction of $1/T_1$ below $T_c$ for the TLL; if we apply the conventional wisdom for exotic superconductivity in two- or three-dimensional correlated electron systems, our findings strongly suggest that the superconducting state is not an isotropic s-wave state.  Instead, an unconventional superconducting ground state with a node in the energy gap seems realized in K$_{2}$Cr$_{3}$As$_{3}$.  This conclusion is consistent with other reports of the possible presence of the nodes in the energy gap based on the measurements on the bulk properties \cite{Bao, Tang, Yuan}. 

Finally, we comment on the nature of the broad side peak B$_{2}$ in Fig.\ 2.  $1/T_{1}T$ at the B$_{2}$ sites is comparable to that of A and B$_{1}$ sites near room temperature, but remains constant, $1/T_{1}T = 0.27$\ s$^{-1}$K$^{-1}$, in the entire temperature range above $T_c$ with no signature of the TLL behavior.  As shown in Fig.\ 3, all the $1/T_{1}$ data points of the B$_2$ sites in the superconducting state are below a naive extrapolation of the corresponding T-linear behavior.  This is consistent with a viewpoint that the B$_2$ sites sense a small energy gap, suggesting the intrinsic nature of the B$_{2}$ sites.  In this scenario, the broad line shape of the B$_{2}$ sites must be a consequence of the defects in their vicinity, which may explain why the TLL behavior is suppressed above $T_c$ and the energy gap is very small below $T_c$.  We cannot rule out, however, an alternative possibility that the B$_{2}$ sites originate from a secondary phase with slightly different K concentration, and the B$_{2}$ peak is merely superposed on the broad tail of the B$_{1}$ sites accounting for $\sim 6$\ \% of the overall intensity.  

To summarize, we demonstrated strong enhancement of spin fluctuations in K$_{2}$Cr$_{3}$As$_{3}$ toward $T_{c}$, obeying a  characteristic power-law predicted theoretically for the TLL.  The absence of the Hebel-Slichter coherence peak of $1/T_1$ just below $T_c$ is followed by a steep decrease, in analogy with unconventional superconductors in higher dimensions with point or line nodes in the energy gap.  K$_{2}$Cr$_{3}$As$_{3}$ exhibits unique quasi 1D properties and belongs to a league of its own, and deserves further attention both as a model system of the TLL and an exotic superconductor.

Acknowledgement: H.\ Z. and T.\ I. thank S.\ Forbes and Y.\ Mozharivskyj for their assistance with their glovebox, and P.\ Dube for SQUID measurements.  The crystal structure in Fig.\ 1 was plotted with Vesta \cite{Vesta}.  The work at McMaster was supported by NSERC and CIFAR.   The work at Zhejiang was supported by the Natural Science Foundation of China (No. 11190023 and No. 11274268) and the National Basic Research Program (No. 2011CBA00103 and No.2014CB921203).\\


\end{document}